\documentclass[12pt]{article}
\usepackage{amsmath}
\usepackage{amssymb}
\usepackage{amsfonts}
\usepackage{nicefrac}
\usepackage[colorinlistoftodos,prependcaption, textsize=small]{todonotes}
\usepackage{setspace}
\usepackage[T1]{fontenc}
\usepackage{times}
\usepackage{palatino} 
\usepackage{lmodern}
\usepackage[latin1]{inputenc}
\usepackage{epsfig}
\usepackage[english]{babel}
\usepackage{color}
\usepackage{graphicx}
\usepackage{dcolumn}
\usepackage{moreverb}
\usepackage{cite}

\usepackage[all,cmtip]{xy}

\numberwithin{equation}{section}

\title{Recurrence relations and general solution of the exceptional Hermite equation}
\author{Alfred Michel Grundland$^1$, Danilo Latini$^2$, Ian Marquette$^{2}$\\ 
{\small $^1$ Centre de Recherches Math\'ematiques, Universit\'e de Montr\'eal }\\ 
{\small Succ. Centre-Ville, CP6128, Montr\'eal (Qc) H3C 3J7 Canada }\\ 
{\small Department of Mathematics and Computer Science, }\\
{\small Universit\'e du Qu\'ebec, CP500, Trois-Rivi\`eres (Qc) G9A 5H7 Canada } \\
{\small $^2$ School of Mathematics and Physics, The University of Queensland,} \\ 
{\small Brisbane, QLD 4072, Australia} 
}

\date{ }
\begin{document}
\baselineskip=22pt plus 1pt minus 1pt
\maketitle

\begin{abstract} 

Exceptional orthogonal Hermite and Laguerre polynomials have been linked to the k-step extension of harmonic and singular oscillators. The exceptional polynomials allow the existence of different supercharges from the Darboux-Crum  and Krein-Adler constructions of supersymmetric quantum mechanics. They also allow the existence of different types of ladder relations and their associated recurrence relations. The existence of such relations is a unique property of these polynomials. Those relations have been used to construct 2D models which are superintegrable, and display an interesting spectrum, degeneracies and finite-dimensional unitary representations. In previous works, only the physical or polynomial part of the spectrum is discussed. It is known that the general solutions are associated with other types of recurrence/ladder relations. We plan to discuss in detail the case of the exceptional Hermite polynomials $X_2^{(1)}$ and to explicitly present new chains obtained by acting with different types of ladder operators. We will exploit a recent result by one of the authors [32], where the general analytic solution was constructed and connected with the non-degenerate confluent Heun equation. The analogue Rodrigues formulas for the general solution are constructed. The set of finite states from which the other states can be obtained algebraically is not unique but the vanishing arrow and diagonal arrow from the diagram of the 2-chain representations can be used to obtain minimal sets. These Rodrigues formulas are then exploited, not only to construct the states, polynomial and non-polynomial, in a purely algebraic way, but also to obtain coefficients from the action of ladder operators also in an algebraic manner based on further commutation relations between monomials of the generators.
\end{abstract}

\\
email: grundlan@crm.umontreal.ca, d.latini@uq.edu.au, i.marquette@uq.edu.au
\bigskip
\noindent
%
%

\newpage

\section{Introduction}

The combination of the Darboux-Crum (DC) and Krein-Adler (KA) transformations \cite{dar88,crum55,kre57,adl94}, also referred to as supersymmetric quantum mechanics \cite{jun96,bag01,and00,fer99}, has been a useful tool in obtaining broader classes of exactly solvable quantum systems. These transformations allow us to obtain the spectra and their almost isospectral Hamiltonians. Their wavefunctions, however, differ greatly and are not necessarily expressed in terms of the hypergeometric class of special functions. \par
%
%
In most of the literature, Darboux-Crum and Krein-Adler transformations (the creating or deleting approach) are concerned with the polynomial (physical) part of the spectrum. It was recognized how those models are associated with various deformations of Heisenberg algebras \cite{and00,fer99} and how physical states are part of the representations which combine finite-and infinite-dimensional representations. Among those models, systems for which the physical part of the spectrum is related to exceptional orthogonal polynomials (EOP) \cite{go10,go09,oda09,cq09} have been the object of particular attention and, in fact, allow the states to be associated with different Rodrigues recurrence relations and patterns of representations of the deformed Heisenberg algebras \cite{fel09,cq11,oda13,ull13,ull14,gom16,gom20}. 
Other special functions are also related such as confluent and Heun equations which themselves have various properties and were studied in a systematic way from point of view of their isomonodromic deformations \cite{der22}. In regard of EOPs, it was demonstrated how the ladder operators are not unique and can even be used to build higher-dimensional models \cite{mar13a,mar13b, mar13c, mar14,car17,dan22} related to the k-step extension of the harmonic and singular oscillators. Those systems are related to multi-indexed Hermite and Laguerre exceptional orthogonal polynomials of type III. A broad review of recent developments of this subject can be found in \cite{go10,go09, oda09,cq09, fel09, cq11, oda13, ull13, ull14, gom16,gom20, mar13a,mar13b, mar13c, mar14, car17}. Those models also play an essential role in the context of superintegrability and the case related to $X_2^{(1)}$ was present in earlier work \cite{gra04} and solved via supersymmetric quantum mechanics (SUSYQM) \cite{fel09,mar09}. The models were also studied before using the Frobenius method \cite{car08,ses10}. Those works have demonstrated how, in fact, one can obtain the polynomial solutions without relying on Darboux transformations. \par
%
%
It was demonstrated how, for isotonic oscillators, the standard ladders are associated with a broader type of representation of the polynomial Heisenberg algebra \cite{mar13a,mar13b,mar13c,mar14}. It was discovered how the polynomial and non-polynomial parts play a role in the study of the quantum systems and their properties and how, in fact, there are generalized states from which different states can be generated \cite{car16}, those states then form wider chains, which we will refer to as 2-chains. Recently \cite{cha20}, the general solutions were presented by a more careful analysis of the gap and series solutions which have been developed. Those solutions connected with confluent Heun equations were applied in the context of the theory of surfaces.  \par
%
%
The purpose of this paper is to develop explicit expressions for the representations and the 2-chains for models related to the Hermite exceptional orthogonal polynomials $X_2^{(1)}$ for the standard ladder, and also for new types of ladders combining both Krein-Adler (KA) and Darboux-Crum (DC) transformations \cite{mar13a,mar13b,mar13c,mar14}. The second objective of this paper is to use the recent formula for the general analytic solutions \cite{cha20} and the connection with the confluent Heun equations related to the exceptional Hermite polynomial $X_2^{(1)}$ to present the action for those states and to demonstrate how they are, in fact, more suitable for the description of those representations of the polynomial Heisenberg algebra. This allows us to present the representations in detail from different perspectives. \par
%
%
The paper is organized as follows. In Section 2, we recall the method for constructing polynomial Heisenberg algebras and how to define states of the 2-chains using the approach from \cite{car16}. We present an explicit action for the standard ladder operators and then, in Section 3, we extend the approach to the non-standard ladder operator involving different types of supercharges. We also point out how a further polynomial Heisenberg algebra can be generated by considering both types of ladder operator. The question of how the algebraic structure can mix different types of ladders has not been addressed in the literature. In Section 4, we use the relation between the EOP Hermite differential equation and the confluent Heun equation to obtain a recurrence relation involving the general solution obtained in \cite{cha20} given by the $\mu_n$ and $\nu_n$ functions expressed in terms of an expression involving generalized hypergeometric functions. These expressions are new. We also rewrite higher-order ladders in terms of first-order ladder operators involving an explicit dependence on the eigenvalues of the Hamiltonian. We also consider the action of $\alpha_n^{(o)}$ and $\alpha_n^{(e)}$ which, combined with $\mu_n$ and $\nu_n$, provide the general solution of the confluent Heun equation where parameters are related to the EOP Hermite ordinary differential equation (ODE). In Section 5, we exploit the connection between the confluent Heun equations and their derivatives. We then obtain the ladders for the Heun equation with specific parameters. Section 6 is devoted to the obtention and modification of a Rodrigues analogue formula in order to present an algebraic construction of the states, polynomial and non-polynomial for $n \in \mathbb{Z}$. This allows us to demonstrate how the states are cyclic i.e. generated by a finite set of states, and how the 2-chain representation and the coefficients from the action of the ladder can also be obtained in a purely algebraic manner. This allows us to specify the form of those coefficients in a more direct way.

\section{Construction of the polynomial Heisenberg algebra and the associated 2-chain representations for the standard ladder}

In this section we recall several aspects of the harmonic oscillator and the deformed oscillator related to the Hermite exceptional orthogonal polynomial $X_2^{(1)}$ \cite{fel09,mar09,mar13a,mar13b,mar13c,mar14,car16} in order to explicitly obtain the action related to 2-chain representations for two different types of ladder operators. As a starting point, consider the simplest Hamiltonian

\[ H_o= - \frac{d^2}{dx^2} + x^2  .\]

This Hamiltonian has been the most studied, possessing the ladder operators $a$ and $a^{\dagger}$

\[ a= \frac{d}{dx} +x ,\quad a^{\dagger}= - \frac{d}{dx} +x \]

which satisfy the Heisenberg algebra

\[ [ a,a^{\dagger}]=2 ,\quad [H_o,a]=-2 a,\quad [H_o,a^{\dagger}]=2 a^{\dagger} .\]

Next, we construct the solution of a time-independent Schr\"{o}dinger problem $H\phi_n = E_n \phi_n$ and obtain the following well-known solution involving the Hermite polynomials $H_n(x)$ of degree $n$
\[ \phi_n = e^{-\frac{x^2}{2}} H_n(x). \]
The associated spectrum is discrete and given by $E_n=2n+1$. We also construct non-normalizable solutions below the ground state in terms of the pseudo-Hermite polynomial
\[ \phi^{-}_n =  e^{\frac{x^2}{2}} H_n(i x) .\]
We label the indices alternatively as  $\phi_{-n-1}=\phi^{-}_n$ and then include those with a negative index with the normalizable solutions and define $\psi_{p}$ for $p=0,\pm 1,\pm 2, ...$. The linearly independent solution is then obtained via 
\[ \tilde{\phi}_{p}= \phi_{p} \int^x \frac{1}{\phi_p^2 (z)} dz \, . \]

The importance of those states was, in particular, pointed out in \cite{car16}. Those states also admit recurrence relations connected with the Rodrigues Formula

\[ a \psi_0 =0 \]
\[ (a^{\dagger})^n \psi_0 \propto \psi_n \, . \]

We consider a more general situation, namely when the Hamiltonian is
\begin{equation}
 H= -\frac{d^2}{dx^2} + x^2 - \frac{16}{(1+2x^2)^2} + \frac{8}{1+2x^2}. 
\end{equation}

This leads to a description of a 1-step extension of the harmonic oscillator with $m=2$, which was studied in \cite{mar13a,mar13b,mar13c,mar14}. In general, for a 1-step extension, there are two types of supercharges, both connecting the Hamiltonian with the harmonic oscillator ( with different numbers of intermediate Hamiltonians ). This ensures the existence of different ladder operators. Those ladder operators lead to a polynomial Heisenberg algebra which has been analyzed from the point of view of differential operator realizations and the action on the physical states $\psi_n$ with $\psi_{-3}$, $\psi_0$, $\psi_1$, $\psi_2$, $\psi_3$, $...$ and the corresponding energies $-3,3,5,7,...$. The spectrum is then given by $E_n=2n+3$ and the corresponding normalizable solutions are

\begin{equation}
  \psi_n \propto \frac{e^{-\frac{1}{2}x^{2}}}{\mathcal{H}_{2}}y_n(x), \quad n=-3, 0, 1, 2,\ldots  
\end{equation}
\begin{equation}
  y_n(x) = \left\{
    \begin{array}{ll}
    1 & {\rm if\ } n=-3, \\[0.2cm]
    - \mathcal{H}_2 H_{n+1} - 4 \mathcal{H}_{1} H_{n} & {\rm if\ } n=0, 1, 2,\ldots,  
    \end{array}
  \right.  \label{EOPhermite}
\end{equation}

where  $\psi_{-3}$ is the fundamental state.

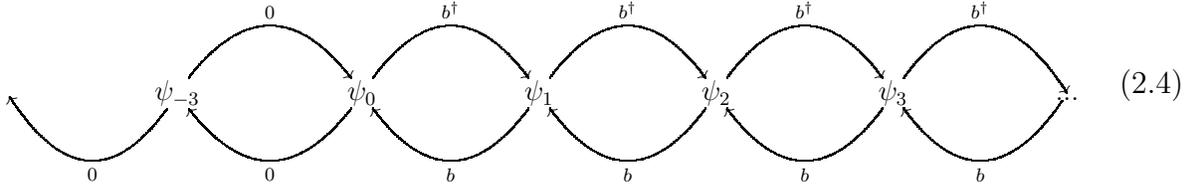
\begin{equation}
\scalebox{0.9}{  \xymatrixcolsep{5pc}\xymatrix@1{
  & \psi_{-3}  \ar@/^{10mm}/[l]^{0}  \ar@/^{10mm}/[r]^{0}  &   \psi_0 \ar@/^{10mm}/[r]^{b^{\dagger}} \ar@/^{10mm}/[l]^{0}   &     \psi_1   \ar@/^{10mm}/[r]^{b^{\dagger}} \ar@/^{10mm}/[l]^{b} &  \psi_2  \ar@/^{10mm}/[l]^{b}   \ar@/^{10mm}/[r]^{b^{\dagger}}   & \psi_3 \ar@/^{10mm}/[r]^{b^{\dagger}}  \ar@/^{10mm}/[l]^{b}  & ... \ar@/^{10mm}/[l]^{b} }}
\end{equation}

Those physical states are related to exceptional orthogonal Hermite polynomials of type III \cite{go10,go09,oda09,cq09,fel09,cq11, oda13,ull13,ull14,gom16,gom20,mar13a,mar13b, mar13c,mar14,car17}. The ladder operators can be constructed from the supercharges

\[ A= \frac{d}{dx} -x - \frac{8x}{2+4x^2} ,\quad A^{\dagger} = - \frac{d}{dx} -x  - \frac{8x}{2+4x^2} \]

using (2.1) and the intertwining relations

\[ A^{\dagger}(H-2) = H_o A^{\dagger} ,\quad (H-2)A=A H_o ,\quad A A^{\dagger} = H+3 ,\quad A^{\dagger} A = H_o +5 .\]

The supercharges $A$ and $A^{\dagger}$ connect the Hamiltonian $H$ with the harmonic oscillator
and provide the ladder operators for the Hamiltonian $H$, which is a deformation of the harmonic oscillator  $b^{\dagger} = A a^{\dagger} A^{\dagger}$ and $b = A a A^{\dagger}$. Here, $b$ and $b^{\dagger}$ satisfy

\begin{equation}
  [ H,b] =-2b ,\quad  [ H,b^{\dagger}]= 2b^{\dagger} 
\end{equation}
\begin{equation}
 b^{\dagger} b = (H-3)(H+1)(H+3) ,\quad b b^{\dagger} = (H-1)(H+3)(H+5) .
\end{equation}

The action on the physical states $\{\psi_{-3},\psi_0,\psi_1,\psi_2,...\}$, (where $\psi_{-3}$ is a singlet as it is annihilated by $b$ and $b^{\dagger}$, and $\psi_0$ is annihilated by $b$) and the action of $b^{\dagger}$ on $\psi_0$ generate an infinite sequence of physical states. Various further identities can be proven and will be useful in using Rodrigues-type formulas and calculating the action of different ladders among those
\begin{equation}
[b,(b^{\dagger})^n ]= (b^{\dagger})^{n-1} ( 6 n H^2  + 4 n (1+3n) H +2 n (-9 + 2n +4 n^2 ) ) 
\end{equation}
\begin{equation}
 [b^{\dagger},b^n] =  b^{n-1} (  -6 n H^2 + 4 n (-7  + 3n ) H -2n (7  - 14 n +4 n^2) ) .
\end{equation}
We will now extend the study of those ladder operators to the action on physical and non-physical states. The general solution contains other linearly independent solutions that we denote $\bar{\psi}_n$ corresponding to all values of the energy ( even the one with no physical states ). Here, we present the two linearly independent solutions obtained via the Darboux transformation from the linearly independent solution of the Harmonic oscillator i.e. we take $\tilde{\psi}_{p}$ as defined previously with $p=0,\pm 1, \pm 2, \pm 3,...$ and we define the state of the deformed oscillator as

\begin{equation}
  \psi_{p}= A \phi_p  ,\quad  \tilde{\psi}_p= A \tilde{\phi}_p .
\end{equation}
However, we exclude $p \neq -3$ ( which we consider because $\psi_{-3}=0$ ), because this state is exactly the zero mode of the supercharge. We need to define the state via $A^{\dagger}\psi_{-3} =0$ and then we add the  linearly independent solution via 

\begin{equation}
 \tilde{\psi}_{-3}= \psi_{-3} \int^x \frac{1}{\psi_{-3}^2 (z)} dz 
\end{equation} 

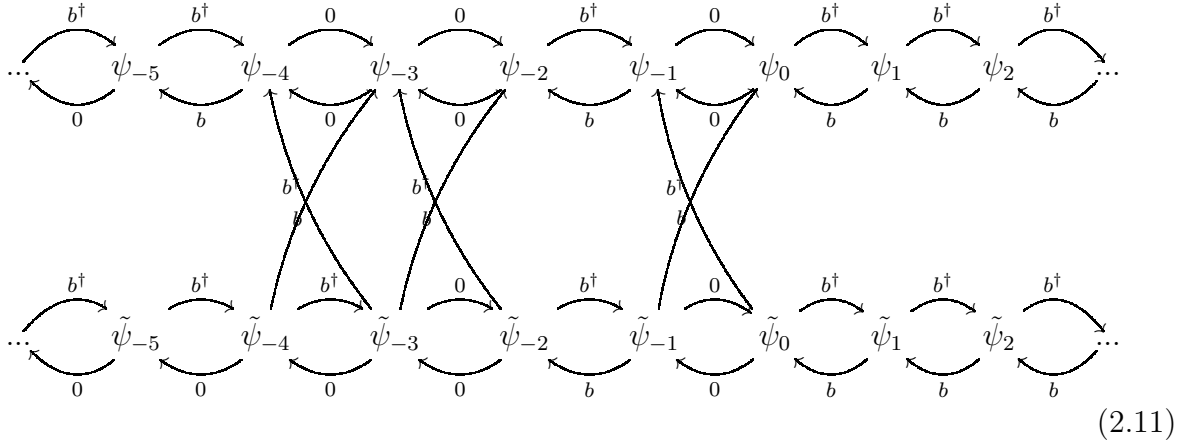
\begin{equation}
\xymatrix{
... \ar@/^{5mm}/[r]^{b^{\dagger}} & \psi_{-5}  \ar@/^{5mm}/[l]^{0}   \ar@/^{5mm}/[r]^{b^{\dagger}}  &  \psi_{-4}   \ar@/^{5mm}/[l]^{b}   \ar@/^{5mm}/[r]^{0} & \psi_{-3}  \ar@/^{5mm}/[l]^{0}   \ar@/^{5mm}/[r]^{0}  &  \psi_{-2} \ar@/^{5mm}/[l]^{0}   \ar@/^{5mm}/[r]^{b^{\dagger}} & \psi_{-1} \ar@/^{5mm}/[l]^{b}   \ar@/^{5mm}/[r]^{0}   &   \psi_{0}   \ar@/^{5mm}/[l]^{0}   \ar@/^{5mm}/[r]^{b^{\dagger}}    &  \psi_{1}   \ar@/^{5mm}/[l]^{b}   \ar@/^{5mm}/[r]^{b^{\dagger}} &  \psi_{2}   \ar@/^{5mm}/[l]^{b}   \ar@/^{5mm}/[r]^{b^{\dagger}} & ... \ar@/^{5mm}/[l]^{b}   \\ \\ \\
...  \ar@/^{5mm}/[r]^{b^{\dagger}} & \tilde{\psi}_{-5} \ar@/^{5mm}/[l]^{0}   \ar@/^{5mm}/[r]^{b^{\dagger}} & \tilde{\psi}_{-4} \ar@/^{3mm}/[uuur]^{b^{\dagger}}  \ar@/^{5mm}/[l]^{0}   \ar@/^{5mm}/[r]^{b^{\dagger}} & \tilde{\psi}_{-3}  \ar@/^{3mm}/[uuur]^{b^{\dagger}} \ar@/^{3mm}/[uuul]^{b}  \ar@/^{5mm}/[l]^{0}   \ar@/^{5mm}/[r]^{0}       & \tilde{\psi}_{-2}\ar@/^{3mm}/[uuul]^{b}  \ar@/^{5mm}/[l]^{0}   \ar@/^{5mm}/[r]^{b^{\dagger}}   &  \tilde{\psi}_{-1}  \ar@/^{3mm}/[uuur]^{b^{\dagger}}   \ar@/^{5mm}/[l]^{b}   \ar@/^{5mm}/[r]^{0}  &    \tilde{\psi}_{0} \ar@/^{3mm}/[uuul]^{b}  \ar@/^{5mm}/[l]^{0}   \ar@/^{5mm}/[r]^{b^{\dagger}} &  \tilde{\psi}_{1}   \ar@/^{5mm}/[l]^{b}   \ar@/^{5mm}/[r]^{b^{\dagger}}    & \tilde{\psi}_{2}   \ar@/^{5mm}/[l]^{b}   \ar@/^{5mm}/[r]^{b^{\dagger}}  & ... \ar@/^{5mm}/[l]^{b}
}
\end{equation}

\begin{equation}
b^\dagger \psi_n = 
\begin{cases}
2(n+3) \psi_{n+1} \hskip 4.5cm n \geq 0\\
2 \imath (n+1)(n+3)(n+4)\psi_{n+1} \hskip 0.5cm -1 \geq n \geq -4\\
4 \imath (n+1)(n+3)\psi_{n+1} \hskip 3.05cm n \leq -5
\end{cases}
\end{equation}

\begin{equation}
b^\dagger \tilde{\psi}_n = 
\begin{cases}
4(n+1)(n+3) \tilde{\psi}_{n+1} \hskip 2.6cm n \geq 0\\
-4 \psi_0 \hskip 5.25cm n=-1\\
-2 \imath \tilde{\psi}_{-1} \hskip 4.9cm n=-2\\ 
-2 \imath \psi_{-2} \hskip 4.9cm n=-3\\
+2 \imath \psi_{-3} \hskip 4.9cm n=-4\\
-2 \imath (n+3) \tilde{\psi}_{n+1} \hskip 3.4cm n \leq -5
\end{cases}
\end{equation}

\begin{equation}
b \,\psi_n = 
\begin{cases}
4n(n+3) \psi_{n-1} \hskip 3.9cm n \geq 0\\
-4 \imath \psi_{-2} \hskip 5.15cm n=-1\\ 
0\hskip 6.275cm n=-2\\
-2 \imath (n+3) \psi_{n-1} \hskip 3.7cm n \leq -3
\end{cases}
\end{equation}

\begin{equation}
b\, \tilde{\psi}_n = 
\begin{cases}
2(n+3) \tilde{\psi}_{n-1} \hskip 3.9 cm n \geq 1\\
6 \psi_{-1} \hskip 5.375cm n=0\\
-8 \imath \tilde{\psi}_{-2} \hskip 4.92cm n=-1\\ 
-8 \imath \psi_{-3} \hskip 4.92cm n=-2\\
- \nicefrac{\imath}{2} \,\psi_{-4} \hskip 4.82cm n=-3\\
4 \imath n(n+3) \tilde{\psi}_{n-1} \hskip 3.6cm n \leq -4
\end{cases}
\end{equation}

\section{ Construction of ladder operators with a Darboux-Crum/Krein-Adler transformation and 2-chain representations}

The other supercharges are:

\[ A_1 = \frac{d}{dx} +x - \frac{1}{x} ,\quad A_1^{\dagger} = - \frac{d}{dx} +x - \frac{1}{x} \]
\[ A_2= \frac{d}{dx} +x + \frac{1}{x} - \frac{8x}{2+4x^2} ,\quad A_2^{\dagger}  =  - \frac{d}{dx} +x + \frac{1}{x} - \frac{8x}{2+4x^2}. \]

The supercharges also satisfy the intertwining relation with H given by (2.1)

\[ (A_2 A_1) H_o = (H+4)(A_2 A_1) ,\quad  H_o (A_1^{\dagger} A_2^{\dagger})= (A_1^{\dagger} A_2^{\dagger}) (H+4) \]

and the factorization

\[ (A_2 A_1)( A_1^{\dagger} A_2^{\dagger}) = (H-1)(H+1) \quad (A_1^{\dagger} A_2^{\dagger}) ( A_2 A_1) = (H_o-5)(H_o-3). \]

We can then form other ladder operators $c$ and $c^{\dagger}$ as follows:

\begin{equation}
 c^{\dagger}= A A_1^{\dagger} A_2^{\dagger} ,\quad c = A_2 A_1 A^{\dagger} 
\end{equation}

and 

\begin{equation}
 [H,c]=-6c ,\quad  [H,c^{\dagger}]= 6 c^{\dagger} 
\end{equation}
\begin{equation}
  c^{\dagger} c =  (H+3)(H-5)(H-7) ,\quad c c^{\dagger} =  (H-1)(H+1)(H+9) .
\end{equation}

\begin{equation}
\scalebox{0.9}{  \xymatrixcolsep{5pc}\xymatrix@1{
 & \psi_{-3} \ar@/^{10mm}/[r]^{c^{\dagger}}    &  \psi_0 \ar@/^{10mm}/[rrr]^{c^{\dagger}} \ar@/^{10mm}/[l]^{c}   &     \psi_1   \ar@/^{10mm}/[rrr]^{c^{\dagger}}  &  \psi_2  \ar@/^{10mm}/[rrr]^{c^{\dagger}}   & \psi_3    \ar@/^{10mm}/[lll]^{c}  & ...  \ar@/^{10mm}/[lll]^{c}  & ... \ar@/^{10mm}/[lll]^{c}  }}
\end{equation}

Here we only include the non vanishing arrows. The action on the polynomial part of the physical states is c, which annihilates $\{\psi_{-3},\psi_1,\psi_2\}$. Therefore,  by acting on those zero modes $c^{\dagger}$ gives three infinite chains of states. We demonstrate the further higher-order commutation relations

\begin{equation}
 [ c, (c^{\dagger})^n]=(c^{\dagger})^{n-1} ( 18 n H^2 + 108 n (n-1) H + 6 n(-1-54 n + 36 n^2)) 
\end{equation}
\begin{equation}
 [c^{\dagger},c^n]= c^{n-1} (  -18 n  H^2 + 108 n (n-1) H - 6 n (-1 -54 n + 36n^2 )) 
\end{equation}

\begin{equation}
\xymatrix{
 ... \ar@/^{5mm}/[rrr]^{0} & \psi_{-5}    \ar@/^{5mm}/[rrr]^{c^{\dagger}}  &  \psi_{-4}    \ar@/^{5mm}/[rrr]^{c^{\dagger}} & \psi_{-3}  \ar@/^{5mm}/[lll]^{0}   \ar@/^{5mm}/[rrr]^{c^{\dagger}}  &  \psi_{-2} \ar@/^{5mm}/[lll]^{c}   \ar@/^{5mm}/[rrr]^{0} & \psi_{-1} \ar@/^{5mm}/[lll]^{c}   \ar@/^{5mm}/[rrr]^{0}   &   \psi_{0}   \ar@/^{5mm}/[lll]^{c}   \ar@/^{5mm}/[rrr]^{c^{\dagger}}    &  \psi_{1}   \ar@/^{5mm}/[lll]^{0}   \ar@/^{5mm}/[rrr]^{c^{\dagger}} &  \psi_{2}   \ar@/^{5mm}/[lll]^{c}    & \psi_{3}   \ar@/^{5mm}/[lll]^{c}    & ... \ar@/^{5mm}/[lll]^{c}   \\ \\ \\
 ...  \ar@/^{3mm}/[uuurrr]^{c^{\dagger}} \ar@/^{5mm}/[rrr]^{0} & \tilde{\psi}_{-5}  \ar@/^{5mm}/[rrr]^{c^{\dagger}} & \tilde{\psi}_{-4}     \ar@/^{5mm}/[rrr]^{c^{\dagger}} & \tilde{\psi}_{-3}  \ar@/^{3mm}/[uuulll]^{c} \ar@/^{5mm}/[lll]^{0}   \ar@/^{5mm}/[rrr]^{c^{\dagger}}       & \tilde{\psi}_{-2}  \ar@/^{5mm}/[lll]^{c}   \ar@/^{3mm}/[uuurrr]^{c^{\dagger}}  \ar@/^{5mm}/[rrr]^{0}   &  \tilde{\psi}_{-1}     \ar@/^{5mm}/[lll]^{c}   \ar@/^{3mm}/[uuurrr]^{c^{\dagger}}  \ar@/^{5mm}/[rrr]^{0}  &    \tilde{\psi}_{0} \ar@/^{5mm}/[lll]^{c}   \ar@/^{5mm}/[rrr]^{c^{\dagger}} &  \tilde{\psi}_{1}  \ar@/^{3mm}/[uuulll]^{c}   \ar@/^{5mm}/[rrr]^{c^{\dagger}}    & \tilde{\psi}_{2}   \ar@/^{3mm}/[uuulll]^{c}   & \tilde{\psi}_{3}   \ar@/^{5mm}/[lll]^{c}    & ... \ar@/^{5mm}/[lll]^{c}
}
\end{equation}

\begin{equation}
c^\dagger \,\psi_n = 
\begin{cases}
-\psi_{n+3} \hskip 5.1cm n \geq 0 \cup \{-3\}\\
0 \hskip 6.1 cm n=-6\\
8 \imath (n+1)(n+2)(n+3)  \psi_{n+3} \hskip 1.3 cm n<0\setminus\{-3,-6\}
\end{cases}
\end{equation}

\begin{equation}
c\,\psi_n = 
\begin{cases}
-8(n-1)(n-2)(n+3) \psi_{n-3} \hskip 1.1cm n\geq 0\\ 
-\imath \frac{n+3}{n}\psi_{n-3} \hskip 4.4cm n< 0
\end{cases}
\end{equation}

\begin{equation}
c^\dagger\,\tilde{\psi}_n = 
\begin{cases}
-8(n+1)(n+2)(n+3) \tilde{\psi}_{n+3} \hskip 1.05cm n\geq 0\\ 
\psi_{2} \hskip 5.85cm n=-1\\
-\imath\psi_{1} \hskip 5.4cm n=-2\\
4\tilde{\psi}_{0} \hskip 5.65cm n=-3\\
-\imath \psi_{-3} \hskip 5.2cm n=-6\\
-\imath \tilde{\psi}_{n+3} \hskip 5 cm n\leq-4 \setminus \{-6\}
\end{cases}
\end{equation}

\begin{equation}
c\,\tilde{\psi}_n = 
\begin{cases}
-\frac{n+3}{n} \tilde{\psi}_{n-3} \hskip 3.4cm n\geq 3\\ 
-\nicefrac{5}{2}\, \psi_{-1} \hskip 3.75cm n=2\\
4 \imath \psi_{-2} \hskip 4.25cm n=1\\
12 \tilde{\psi}_{-3}\hskip 4.155cm n=0\\
\nicefrac{\imath}{12} \,\psi_{-6}  \hskip 3.95cm n=-3\\
8 \imath \bigl(n(n^2-7)+6\bigl)\tilde{\psi}_{n-3}  \hskip 1.2cm n<0 \setminus \{-3\}
\end{cases}
\end{equation}

We can show using a direct approach that only two types of ladders of order 3 exist (the lowest order ). A third type of fifth-order differential operator can also be constructed. In general, different types of ladder operators can exist for a k-step extension of the harmonic oscillator but no classification has been performed so far. When the potential is unknown, the existence of ladder operators is, in general, a nonlinear problem, but when the potential is known, the compatibility equations are all linear and can be solved explicitly. \par 
%
%
The two ladders of minimal order can be combined in a common algebraic structure. In view of the differential operators, one can demonstrate the following relations

\[ (b^3)c^{\dagger} = -H^6 -24 H^5 -205 H^4  -720 H^3 - 739 H^2 + 744 H + 945 \]
\[ c(b^{\dagger})^3 = -H^6 + 12 H^5 -25 H^4  -120 H^3 + 34 H^2 + 108 H -315 . \]

Then, defining $B=b^3$ and $B^{\dagger} = (b^{\dagger})^3$, we are able to close the relations as an operator algebra.

\section{ Connection with the confluent Heun equations and the general solution  }

In this section we apply the construction of the ladder in the context of the EOP Hermite ODE in a different form. Specifically, we transform the second-order ODE used in \cite{cha20} into the confluent Heun equation with particular values of the parameters. In order to connect the Schr\"{o}dinger equation with the confluent Heun equation, we consider the gauge/similarity ($\psi_{-3}^{-1} (H+3) \psi_{-3}$) and then obtain the second order ODE

\[ ( \frac{d^2}{dx^2} - ( 2 x + \frac{8 x}{1+2x^2} ) \frac{d}{dx} + 2n ) f_n(x) =0 .\]

The gauge/similarity transformation preserves the spectrum, the existence of the ladder, the commutation relations and the polynomial Heisenberg algebra generated by the ladder operators. The wavefunctions are only related via $\psi_n = f_n \psi_{-3}$. We denote

\[ \tilde{b} = - \frac{d^3}{dx^3} + ( 2 x + \frac{12 x}{1+2x^2} ) \frac{d^2}{dx^2} + ( -4 + \frac{24}{(1+2x^2)^2} - \frac{16}{1+2x^2} ) \frac{d}{dx}. \]

Similarly we get

\[ \tilde{c} =- \frac{d^3}{dx^3} + \frac{12x}{1+2x^2} \frac{d^2}{dx^2} -  \frac{48x^2}{(1+2x^2)^2} \frac{d}{dx} . \]

The raising operators can be obtained directly from the lowering in both cases.

\subsection{ First-order form of the ladder operators }

In most of the literature on ladder operators of exceptional orthogonal polynomials, the ladders are expressed only in terms of higher-order differential operators.
As pointed out in the context of SUSYQM and of different shape-invariant potentials, other alternative forms can be convenient and written as first-order ladder operators which depend on the Hamiltonian eigenvalues. Here we will present an alternative form by using the ODE

\[ f''_n = -2 nf_n + ( 2 x + \frac{8x}{1+2x^2} ) f'_n  \]

and the equation obtained by taking the derivative. This allows us  to rewrite $\tilde{c}$, $\tilde{c}^{\dagger}$,  $\tilde{b}$, $\tilde{b}^{\dagger}$ as first-order operators depending explicitly on $n$

\[ \tilde{b}_n = - \frac{8 n x}{(1+2x^2)} + \frac{2(-3 +n + 4 n x^2 + 4 (n-1) x^4)}{(1+2x^2)^2} \frac{d}{dx} \]

\[ \tilde{c}_n= \frac{4n x(-1+4x^4)}{(1+2x^2)^2}+ \frac{2(-5 +n +2(-5 +2n)x^2 +4(-5 +n)x^4 -8 x^6)}{(1+2x^2)^2} \frac{d}{dx}. \]

\subsection{ The general solutions $\nu_n(x)$ and $\mu_n(x)$ }

In this section, our discussion will facilitate the computation of the action on $\mu_n$ and $\nu_n$. We have

\[ \nu_n(x) = x-\frac{1}{3}(n-5)x^3 + \sum_{k=3}^{\infty} 2^{k-1} ( (-1)^{k+1} \frac{(n-((2(k-1))^2+1))}{(2k-1)!} \prod_{j=1}^{k-2} ( n-2(1+j)+1) x^{2k-1}) \]
\[ \mu_n(x) =1- n x^2 + \frac{n(n-10)}{6} x^4 + \sum_{k=3}^{\infty} 2^k ( (-1)^k \frac{n(n-((2k-1)^2 +1))}{(2k)!} \prod_{j=1}^{k-2} ( n-2(1+j) ) x^{2k} ) .\]

These expressions admit representations as generalized hypergeometric functions which can facilitate the action and consequently can be expressed as follows :

\[ \mu_n(x) =1 -n x^2 + \frac{1}{6} n(n-10) x^4 -\frac{1}{6(n-2)}\biggl(-12 + 6n+12n x^2 - 6 n^2 x^2+ 20 n x^4-12 n^2 x^4 +n^3 x^4  \]
\[+ 24 n x^2 {_1F_1}\left (1-\frac{n}{2},\frac{3}{2},x^2 \right )-6(n-2)\, {_1F_1}\left (-\frac{n}{2},\frac{1}{2},x^2\right )-24 n x^2 {_2}F_2\left ( \{2,1-\frac{n}{2}\},\{1,\frac{3}{2}\},x^2\right ) \biggl)\]
\[=   {_1F_1}\left (-\frac{n}{2},\frac{1}{2},x^2 \right ) -\frac{4}{3}n x^4 {_1F_1}\left (2-\frac{n}{2},\frac{5}{2},x^2 \right )\]

\[\nu_n(x) =x-\frac{1}{3}(n-5)x^3 + \frac{1}{3(n-1)x} \biggl( 3 x^2 -3 n x^2 + 5 x^4 -6 n x^4 + n^2 x^4 -15 x^2 {_1F_1}\left( \frac{1}{2}-\frac{n}{2},\frac{3}{2},x^2\right) \]
\[ + 3 n x^2 {_1}F_1\left (\frac{1}{2}-\frac{n}{2},\frac{3}{2},x^2\right) + 24 x^2 {_2F_2}\left ( \{2,\frac{1}{2}-\frac{n}{2}\},\{1,\frac{3}{2}\},x^2\right ) -12 x^2 {_3F_3}\left ( \{2,2,\frac{1}{2}-\frac{n}{2}\},\{1,1,\frac{3}{2}\},x^2\right ) \biggl)\]
\[ = (x+2x^3) {_1F_1}\left (\frac{1}{2}-\frac{n}{2},\frac{3}{2},x^2 \right ) + \frac{2}{3} x^3 (-1+ 2x^2){_1F_1}\left (\frac{3}{2}-\frac{n}{2},\frac{5}{2},x^2 \right ) .\]

Generalized hypergeometric functions admit various recurrence relations and formulas. They can be used to calculate the action of $b_n$, $b_n^{\dagger}$, $c_n$, $c_n^{\dagger}$ on $\nu_n(x)$ and $\mu_n(x)$

\[ \tilde{b}_n \mu_n = (4 n -4n^2 ) \nu_{n-1},\quad \tilde{b}_n \nu_n = (-6 +2n) \mu_{n-1} \]

and consequently the differential operators independent of $n$ and denoted $\tilde{b}$ can be evaluated. The zero modes are, respectively, $\mu_0$, $\mu_1$ and $\nu_3$. For the creation (raising) operators, we have the following recurrence relations

\[ \tilde{b}_n^{\dagger} \mu_n = (4 n +4n^2 ) \nu_{n+1},\quad \tilde{b}_n^{\dagger} \nu_n = (4-2n) \mu_{n+1} . \]

Also, from the differential operator $\tilde{b}^{\dagger}$ independent of $n$, we obtain the same recurrence relations. The zero modes are $\mu_0$, $\mu_{-1}$ and $\nu_2$. Also, we obtain the following recurrence relations from the other set of ladder operators 

\[ \tilde{c}_n \mu_n = (16 n -4n^2 ) \nu_{n-3}, \quad \tilde{c}_n \nu_n = (-10 +2n) \mu_{n-3}, \]
\[ \tilde{c}_n^{\dagger} \mu_n = 4(n-1)(n+3) \nu_{n+3}, \quad \tilde{c}_n^{\dagger} \nu_n = (4-2n) \mu_{n+3} .\]

Those results demonstrate how most of the general solutions of the EOP Hermite ODE have recurrence relations in a very simple form. Those functions $\mu_n$ and $\nu_n$
are, however, only a part of the general solution. In the next section we will add the other component of the general solution and this will explain how even when the general solution is written in this way, the action of the ladder still has an involved form.

\subsection{The coefficients $\alpha_n^{(o)}$ and $\alpha_n^{(e)}$}

In order to define the other linearly independent solutions of the EOP Hermite ODE we need to use the expression 

\[ \alpha_n(z) = M_{3}(n) H_n(z) \]
\[ H_n(z)=M_1(n) \nu_n (z) ,\quad \forall \quad n=2l-1, l \geq 2 \]
\[ H_n(z)=M_2(n) \mu_n(z) ,\quad \forall \quad n=2l, l \in \{0,2,3,4,...\}  \]

with

\[ M_3(n) = \begin{cases} 
 -2  & n=1 \\
   2  & n=2 \\
	-M_1^{-1}(n) & n \in 2 \mathbb{N}-1 \setminus \{1\} \\
   M_2^{-1}(n) & n \in 2 \mathbb{N} \setminus \{2\} 
													\end{cases}  \]
													
and													

\[ M_1(n)= \frac{(-1)^{\frac{n+1}{2}} n! 2^{\frac{(n+1)}{2}} }{ P_{(1,1)}(n) \prod_{j=1}^{\frac{(n-3)}{2}} (n -2(j+1)+1)} \]
\[ M_2(n) =(-1)^{\frac{(n+2)}{2}} 2^{n-1} \pi^{-\frac{1}{2}} \Gamma(\frac{n-1}{2}), \quad \forall n \in 2 \mathbb{N} \setminus\{2\} \]

where $P_{(1,1)}(n) = (n-1)(n-2)$. We also have

\[   \alpha_1^{(o)}=-8 x - 2 e^{x^2} \sqrt{\pi}(1-2x^2)(1-Erf(x)) = (-2 \sqrt{\pi} \mu_1 -4  \nu_1 ),    \]
\[  \alpha_2^{(e)}= 4+ 8 x^2 + 8 \sqrt{\pi} x e^{x^2} (1-Erf(x))= \alpha_2^{(e)}=4(\mu_2 + 2 \sqrt{\pi} \nu_2 ). \]

There are further relations such as

\[ \alpha_n^{(o)}=-\nu_n, \quad  n=3,5,7,... \]
\[ \alpha_n^{(e)}=\mu_n,\quad n=4,6,8,... \, .\]

We evaluate  the action of $b_n$, $b_n^{\dagger}$, $c_n$, $c_n^{\dagger}$ and $\tilde{b}$, $\tilde{b}^{\dagger}$, $\tilde{c}$, $\tilde{c}^{\dagger}$, and consider the $\alpha_n^{(o)}$ for odd indices and $\alpha_n^{(e)}$ for even indices

\[ \tilde{b}\alpha_1^{(o)}  = 16 \alpha_0^{(e)} ,\quad \tilde{b}^{\dagger} \alpha_1^{(o)} =-2 \alpha_2^{(e)}\]

with $\alpha_0 := \mu_0\equiv \alpha_0^{(e)}\equiv \alpha_0^{(o)}=1  $. We will only list the relations for other ladder operators

\[ \tilde{b} \alpha_2^{(e)} = 8 \alpha_1^{(o)}\quad \tilde{b}^{\dagger} \alpha_2^{(e)} = -96  \alpha_3^{(o)} \]
\[ \tilde{b} \alpha_{n}^{(o)} = ( 6 -2n) \alpha^{(e)}_{n-1} ,\quad n=3,5,7,...\]
\[ \tilde{b}^{\dagger} \alpha_{n}^{(o)} = (2n-4) \alpha_{n+1}^{(e)},\quad n=3,5,7,...\]
\[ \tilde{b} \alpha_{n}^{(e)} = 4n(n-1) \alpha^{(o)}_{n-1},\quad n=4,6,8,... \]
\[\tilde{b}^{\dagger} \alpha_{n}^{(e)} = -4n(n+1)\alpha^{(o)}_{n+1},\quad n=4,6,8,... \]
\[\tilde{c} \alpha_1^{(o)} = 32 \mu_{-2}  - 24 \sqrt{\pi} \nu_{-2} ,\quad \tilde{c}^{\dagger} \alpha_1^{(o)} = - 8 \alpha^{(e)}_4  ,\quad \tilde{c} \alpha_2^{(e)} = - 48 \sqrt{\pi} \mu_{-1}  + 64  \nu_{-1}\]
\[ \tilde{c}^{\dagger} \alpha_2^{(e)} = -80  \alpha^{(o)}_5 ,\quad \tilde{c} \alpha_{n}^{(o)} = \begin{cases} 4 \alpha_0\hskip 2.75cm n=3\\ 
	(10-2n)\alpha_{n-3}^{(e)} \quad n=5,7,9,...\end{cases} ,\quad \tilde{c}^{\dagger} \alpha_{n}^{(o)} = (2n-4) \alpha^{(e)}_{n+3}  \quad n=3,5,...\]
\[ \tilde{c} \alpha_{n}^{(e)} =  4n(n-4)\alpha^{(o)}_{n-3},\quad n=4,6,...  \]
\[ \tilde{c}^{\dagger} \alpha_{n}^{(e)} = 4(1-n)(n+3)\alpha^{(o)}_{n+3},\quad n=4,6,... \, . \]

In this context, the coefficient of the action of the ladder differs greatly from the action in Sections 2 and 3 which were obtained via Darboux transformation of the eigenfunctions of the harmonic oscillator for the two linearly independent solutions. Here we used series solution and connection with generalized and hypergeometric functions. The coefficients are now connected with the functions obtained by \cite{cha20}, which constitute the general solution.

\section{ Ladders and representations in the Heun form }

The purpose of this section is to connect the obtained results with the confluent Heun and Heun equations. The EOP $X_{2}^{(1)}$ Hermite ODE 

\[ \omega'' - 2( x + \frac{4x}{1+2x^2} ) \omega' + 2n \omega =0 \]

can be written in the following form in terms of the variable $s$  ($y= -2x^2$) \cite{cha20}

\[ \frac{d^2 \omega}{ds^2} + \left ( \frac{\gamma}{s} + \frac{\delta}{s-1} + \epsilon \right ) \frac{d\omega}{ds} + \left (\frac{\alpha s -q}{s(s-1)}\right ) \omega =0 \]

where the parameters of the confluent Heun equation take the following values : $\gamma=\frac{1}{2}$, $\alpha =q= -\frac{n}{4}$, $\epsilon = \frac{1}{2}$, $\delta =-2$. 
Similarly, the ladder operators can be written in terms of the variable $s$ :

\[ \bar{b} =-16 s\sqrt{-2s}\partial_s^3-\frac{8\sqrt{-2s}(s^2-4s-3)}{s-1} \partial_s^2+\frac{4\sqrt{-2s}(s^2-4s-9)}{(s-1)^2} \partial_s\]
\[\bar{c} = -16  s \sqrt{-2s}\partial_s^3 -  \frac{24 \sqrt{2} s (s+1)}{\sqrt{-s}(s-1)}  \partial_s^2 - \frac{24  \sqrt{-2s} (s+1)}{(s-1)^2}  \partial_s.\]
The functions $\mu$ and $\nu$ can also be written in term of $s$ and expressed in term of the confluent hypergeometric function ${_1F_1}$ only 

\[ \bar{\nu}_n(s)= \frac{\sqrt{-s}}{3\sqrt{2}} \left (-3 (s-1) \,{_1F_1}\left (\frac{1}{2}-\frac{n}{2},\frac{3}{2},-\frac{s}{2}\right ) +s(s+1)\, {_1F_1}\left (\frac{3}{2}-\frac{n}{2},\frac{5}{2},-\frac{s}{2}\right ) \right ) \]
\[\bar{\mu}_n(s)= \frac{1}{n-2} \left ( -(s^2+(n+1)s-n+2)\, \,{_1F_1}\left (-\frac{n}{2},\frac{1}{2},-\frac{s}{2}\right ) +(1+n)s(s+1)\, {_1F_1}\left (-\frac{n}{2},\frac{3}{2},-\frac{s}{2}\right ) \right ) \]
with recurrence relations

\[ \bar{b} \bar{\mu}_n = 4n (1-n) \bar{\nu}_{n-1} ,\quad \bar{b} \bar{\nu}_n = ( 2n-6) \bar{\mu}_{n-1} \]
\[ \bar{b}^{\dagger} \bar{\mu}_n =4n (n+1) \bar{\nu}_{n+1} ,\quad \bar{b}^{\dagger} \bar{\nu}_n =(4-2n) \bar{\mu}_{n+1} \]
\[ \bar{c} \bar{\mu}_n = 4 n(4-n) \bar{\nu}_{n-3} ,\quad \bar{c} \bar{\nu}_n = (2n-10) \bar{\mu}_{n-3} \]
\[ \bar{c}^{\dagger} \bar{\mu}_n =4(n-1)(n+3) \bar{\nu}_{n+3} ,\quad \bar{c}^{\dagger} \bar{\nu}_n =(4-2n) \bar{\mu}_{n+3} .\]

We now want to look at the ladder operators for the derivative of $\omega$. The connection between the derivative of a Heun function was discussed in a recent work \cite{cha20}. However recurrence relations for derivatives of the Heun polynomial appear to be an unexplored problem in the literature. Setting $v(s)= d\omega/ds$, we then perform a change of variable $s=\frac{s_1}{s_1 -1}$ and $v(s_1) =(1-s_1)^{(1-\frac{n}{2})} W(s_1) =g_n W(s_1)$. This can also be done in operator form as the gauge transformation $\hat{H}_1 = g_n^{-1} \bar{H}_1 g_n$ with $\hat{H}_1$ acting on $W(s_1)$ as  $\hat{H}_1 W(s_1)=0$, which gives

\[ (-1+s_1)^2 s_1 \frac{d^2}{ds_1^2}W + \frac{1}{4} (6 +4 s_1 (-8 +n + 6 s_1 -n s_1)) \frac{d}{ds_1} W  + \frac{1}{4} (n-4) (4 + (n-6) s_1 ) W =0. \]

Here, $W(s_1)$ has the linearly independent solutions

\begin{align*}
		\hat{\mu}_n(s_1)&=\frac{n (1-s_1)^{\frac{n-6}{2}}}{6 (n-2)} \biggl(3 (s_1(1-n)+n-2) \, _1F_1\left(1-\frac{n}{2};\frac{3}{2};\frac{s_1}{2-2 s_1}\right)\\
&+(n+1) s_1\, {_1}F_1\left(1-\frac{n}{2};\frac{5}{2};\frac{s_1}{2-2 s_1}\right)\biggl)\\
		\hat{\nu}_n(s_1)&=\frac{ \sqrt{\frac{1}{2-2 s_1}} (1-s_1)^{\frac{n-6}{2}}}{6 \sqrt{s_1} (n-1)}\biggl(3 \left(\left(s_1^2-1\right) n+1\right) \, _1F_1\left(\frac{1}{2}-\frac{n}{2};\frac{3}{2};\frac{s_1}{2-2 s_1}\right)
\\		
&-s_1 (n+2) ((s_1-1) n+1) \, _1F_1\left(\frac{1}{2}-\frac{n}{2};\frac{5}{2};\frac{s_1}{2-2 s_1}\right)\biggl) .
		\end{align*}

We want to obtain the corresponding ladder operator from the same transformation i.e. taking the derivative and the above gauge transformation. In particular we consider $\frac{d}{ds} \bar{b}^{\dagger}$ and $\frac{d}{ds} \bar{b}$ and $v=\frac{d\omega}{ds}$. Here we only present the ladder operators $b$ and $b^{\dagger}$ acting on $\hat{\mu}$ and $\hat{\nu}$ in the $s_1$ variable. Explicitly, the lowering operator becomes

\[ \hat{b}_n = -16 \sqrt{2} (-1+s_1)^4 s_1^{\frac{3}{2}} \frac{d^3}{ds_1^3} + ( 8 \sqrt{2} (-1+s_1)^2 \sqrt{s_1} (-6 +(31-3n)s_1 +3 (-8+n) s_1^2) \frac{d^2}{ds_1^2} \]
\[ - \frac{1}{\sqrt{s_1}} (4 \sqrt{2} (-1+s_1) (-3 +s_1 (81 -12n -(-6+n) ( -38 +3n)s_1 +3(-8 +n)(-6+n) s_1^2 )) \frac{d}{ds_1} \]
\[ - \frac{1}{\sqrt{s_1}}(2 \sqrt{2} (3(-5+n) +3(-4+n)(-15+2n)s_1+ (-15+n)(-8+n)(-4+n)s_1^2\]
\[ -(-8+n)(-6+n)(-4+n)s_1^3)) \]

and we have

\[\hat{b}_n \hat{\mu}_n = 4n(1-n) \hat{\nu}_{n-1} ,\quad \hat{b}_n \hat{\nu}_n =(2n-6)\hat{\mu}_{n-1}\]
\[\hat{b}^{\dagger}_n \hat{\mu}_n =4n (n+1) \hat{\nu}_{n+1} ,\quad \hat{b}^{\dagger}_n \hat{\nu}_n =(2n-4) \hat{\mu}_{n+1} .\]

\section{Rodrigues-type formula for both third ladder operators}

We consider the problem of constructing algebraically all the solutions of the differential equation for the exceptional Hermite polynomial $X_2^{(1)}$. It was demonstrated how, from the ladder operators $\{b^{\dagger},b\}$, $\{c^{\dagger},c\}$, one can generate in a purely algebraic way the physical states for $\{\psi_n, n=-3,0,1,2,...\}$. These can be viewed as analogous to the Rodrigues formula

\[ b \psi_{-3} =0 ,\quad b^{\dagger} \psi_{-3} =0 \]
\[ b \psi_0 =0  ,\quad \psi_{n} = (b^{\dagger})^n \psi_0 .\]

We then need two initial zero modes to induce all the physical states. Also we can then reduce the problem of finding the action of different operators to an algebraic
setting. We obtain directly

\[  b^{\dagger} \psi_{n} = \psi_{n+1}  ,\quad b \psi_{n} = ( 8 n (n+2)(n+3) ) \psi_{n-1} .\]

Interestingly, there exists another construction and it can also be put completely into an algebraic setting

\[ c \psi_{-3} =0 ,\quad c \psi_1 =0 ,\quad c \psi_2 =0 \]

and

\[ \psi_{3n-3}=(c^{\dagger})^n \psi_{-3}  ,\quad \psi_{3n+1}=(c^{\dagger})^n \psi_{1}  ,\quad \psi_{3n+2}=(c^{\dagger})^n \psi_{2} \]

and the action is then given by

\[  c^{\dagger} \psi_{3n-3} = \psi_{3(n+1)-3}  ,\quad  c^{\dagger} \psi_{3n+1} = \psi_{3(n+1)+1}  ,\quad  c^{\dagger} \psi_{3n+2} = \psi_{3(n+1)+2} \]

\[  c \psi_{3n-3} = 24 n (3n-5)(3n-4)  \psi_{3(n-1)-3}  ,\quad  c \psi_{3n+1} = 24 n (3n-1)(3n+4) \psi_{3(n-1)+1},\]
\[  c \psi_{3n+2} = 24 n (3n+1)(3n+5) \psi_{3(n-1)+2} . \]

The action is not described for either of the cases of the gap states i.e. $n=-2,-1$, the unbounded states with $n \leq -4$ or the other solutions of the Hamiltonian denoted $\bar{\psi}_n$ for $ n \in \mathbb{Z}$. A recent work by one of the authors \cite{fil20} discussed the idea of a recurrence relation connecting $\psi_{-2}$, $\psi_{-1}$ with the other physical states. Here, we will discuss this connection using another algebraic approach. We consider the case $n=\{-3,-2,-1,0,1,...\}$, $\psi_n$. Using ladder operators of type $b$, we obtain $\psi_{-3}$, $\psi_{-2}$ and $\psi_0$ as initial states which induce all other states

\[ \psi_{-1} = (b^{\dagger}) \psi_{-2}  ,\quad \psi_n = (b^{\dagger})^n \psi_0 . \]

In this case, we want to induce other $\psi_n$ with $n \leq -4$ using

\[ \psi_{-4-n} = (b)^n \psi_{-4} .\]

We can then obtain the action as

\[ b^{\dagger} \psi_{-1}=0,\quad b \psi_{-1}=-4 i \psi_{-2}, \quad b \psi_{-2}=0 . \]

Then

\[  b^{\dagger} \psi_n = \psi_{n+1}  ,\quad b \psi_n = 8 n (n+2)(n+3) \psi_{n-1} \]

\[ b \psi_{-4-n} = \psi_{-4-(n+1)}  ,\quad b^{\dagger} \psi_{-4-n} = - 8 n (n+1)(n+3) \psi_{-4-(n-1)} . \]

For $\psi_n$ with $n \in \mathbb{Z}$, we can use the $c$-type ladder operators as well as the initial state to generate all the states. The initial states are $\{\psi_{-6},\psi_{-1},\psi_{-2}\}$

\[ \psi_{-3n-6} = ( c)^{n} \psi_{-6}  ,\quad \psi_{3n-6} = (c^{\dagger})^n \psi_{-6}  ,\quad \psi_{-1-3n} = (c)^n \psi_{-1} \]
\[ \psi_{-1+3n} = (c^{\dagger})^n \psi_{-1}  ,\quad \psi_{-2+3n}= (c^{\dagger})^n \psi_{-2}  ,\quad \psi_{-2-3n}= (c)^n \psi_{-2} \]

and then the action will be given by the formula

\[ c \psi_{3n-6} = 24 (n-1) (3 n-8) (  3 n-7) \psi_{3(n-1)-6}  ,\quad  c^{\dagger} \psi_{3n-6}=   \psi_{3(n+1)-6} \]
\[ c \psi_{3n-1} =24 (n-1) (3 n-2) (3 n+2) \psi_{3(n-1)-1}  ,\quad  c^{\dagger} \psi_{3n-1}=   \psi_{3(n+1)-1} \]
\[ c \psi_{3n-2} = 24 (n-1) ( 3 n-4) (3 n+1)\psi_{3(n-1)-2}  ,\quad  c^{\dagger} \psi_{3n-2}=   \psi_{3(n+1)-2} \]

and

\[ c^{\dagger} \psi_{-3n-6} = -24 n (4+3n)(5+3n) \psi_{-3(n-1)-6}  ,\quad  c\psi_{-3n-6}=   \psi_{-3(n+1)-6} \]
\[ c^{\dagger} \psi_{-3n-1} = -24 n (3n-5)(3n-1) \psi_{-3(n-1)-1}  ,\quad  c \psi_{-3n-1}=   \psi_{-3(n+1)-1} \]
\[ c^{\dagger} \psi_{-3n-2} = -24 n (3n-4)(3n+1) \psi_{-3(n-1)-2}  ,\quad  c \psi_{-3n-2}=   \psi_{-3(n+1)-2} \,.\]

Now we turn to the problem of inducing all solutions of the Hermite equation of $X_2^{(1)}$, $\psi_n$ and $\bar{\psi}_n$. Those have been explictly calculated in terms of a series for $n \geq -3$. Here we will construct $\psi_n$ and $\bar{\psi}_n$ $\forall n \in \mathbb{Z}$ using an algebraic approach. For the $b$-type ladder operators we will need four states $\bar{\psi}_{-2}$, $\psi_{-2}$, $\bar{\psi}_{-4}$ and $\psi_{0}$

\[ \psi_{-3} = b \bar{\psi}_{-2} ,\quad \bar{\psi}_{-1} =b^{\dagger} \bar{\psi}_{-2}  ,\quad \psi_{n-2} = (b^{\dagger})^{n+1} \bar{\psi}_{-2},\quad n\geq 1\]
\[ \psi_{-1} = b^{\dagger} \psi_{-2}  ,\quad \bar{\psi}_{-4-n} = (b)^n \bar{\psi}_{-4} ,\quad \bar{\psi}_{-3} =b^{\dagger} \bar{\psi}_{-4} \]
\[ \psi_{-4-n} = (b)^n b^{\dagger} \bar{\psi}_{-4}  ,\quad \bar{\psi}_n = (b^{\dagger})^n \bar{\psi}_0  ,\quad b^{\dagger} \bar{\psi}_n = \bar{\psi}_{n+1} \]
\[ b \bar{\psi}_n = 8n(2+n) (3+n) \bar{\psi}_{n-1} \, .\]

From the above relations together with the others given previously we get

\[ b \psi_{n-1}=8 (n-1)(n+1)(n+2) \psi_{n-2}  ,\quad b^{\dagger} \psi_{n-2} = \psi_{n-1} \]
\[ b  \bar{\psi}_{-4-n}= \bar{\psi}_{-4-(n+1)}  ,\quad b^{\dagger} \bar{\psi}_{-4-n} = -4n (6+n)(1+2n) \bar{\psi}_{-4-(n-1)} + \psi_{-4-n} \]
\[ b^{\dagger} \psi_{-4-n} = - 8 n (n+1)(n+3)\psi_{-4-(n-1)}  ,\quad b \psi_{-4-n} = \psi_{-4-(n+1)} \, .\]
We will now use the $c$-type ladder operators and the functions $\bar{\psi}_1$, $\bar{\psi}_2$, $\bar{\psi}_{-3}$, $\psi_{-3}$, $\psi_{-6}$. A pattern is
obtained from the diagonal arrow and the one vanishing in the 2-chain representation obtained by the action on the states constructed using ladders and supercharges

\[ \bar{\psi}_{1-3n} = (c^{\dagger})^n \bar{\psi}_1 ,\quad n \geq 0  ,\qquad \psi_{3n-2} = (c^{\dagger})^n b \bar{\psi}_1,\quad n \geq 0 \]
\[ \psi_{1-3n} = (c)^{n} \bar{\psi}_1 ,\quad n \geq 0  ,\qquad \bar{\psi}_{1+3n} = (c^{\dagger})^{n} \bar{\psi}_1 ,\quad n \geq 0 \]
\[ \bar{\psi}_{2+3n} = (c^{\dagger})^n \bar{\psi}_2  ,\quad \psi_{-1+3n} = (c^{\dagger})^n c \bar{\psi}_2  ,\quad \psi_{2-3n} = (c)^{n} \bar{\psi}_2  ,\quad \bar{\psi}_{-3n-3} = (c)^n \bar{\psi}_{-3} \]
\[ \bar{\psi}_{3n-3} = (c^{\dagger})^n \bar{\psi}_{-3}  ,\quad \psi_{3n-3} = (c^{\dagger})^n \psi_{-3}  ,\quad \psi_{-3n-6} = (c)^n \psi_{-6} \] 

and we get the following 

\[ c \bar{\psi}_{1-3n} = 24n (3n-1)(4+3n) \bar{\psi}_{1-3(n-1)} + \psi_{3n-2}   ,\quad c \psi_{3n-2} =   24n (3n-1)(4+3n) \bar{\psi}_{3(n-1)-2} + \psi_{3n-2}    \]
\[ c \psi_{1-3n} = \psi_{1-3(n+1)}   ,\quad  c \bar{\psi}_{2+3n} =  24 n (1 + 3 n) (5 + 3 n)\bar{\psi}_{2+3(n-1)} + \psi_{-1+3n}  \]
\[ c \psi_{-1+3n} = 24(n-1)(3n-2)(3n+2)\psi_{-1+3(n-1)}   \]
\[ c \psi_{2-3n} = \psi_{2-3(n+1)}  ,\quad c \bar{\psi}_{-3n-3} = \bar{\psi}_{-3(n+1)}  \]
\[ c \bar{\psi}_{3n-3} = 24 n (-5 + 3 n) (-4 + 3 n) \bar{\psi}_{3(n-1)-3}   ,\quad c \psi_{3n-3} = 24 n (-5 + 3 n) (-4 + 3 n) \psi_{3(n-1)-3}  \]
\[ c \psi_{3n-6} = \psi_{3(n+1)-6}  \]

and

\[ c^{\dagger} \bar{\psi}_{1-3n} = \bar{\psi}_{1-3(n+1)} ,\quad c^{\dagger} \psi_{3n-2} = \psi_{3(n+1)-2} \]
\[ c^{\dagger} \psi_{1-3n} = -24(n-1)(3n-7)(3n-2)\psi_{1-3(n-1)} ,\quad c^{\dagger} \psi_{1+3n} = \psi_{1+3(n+1)}     \]
\[ c^{\dagger} \bar{\psi}_{2+3n} = \bar{\psi}_{2+3(n+1)}  ,\quad c^{\dagger} \psi_{-1+3n} = \psi_{-1+3(n+1)} \]
\[ c^{\dagger} \psi_{2-3n}  = (768  -24 n (68 - 45 n + 9 n^2)) c^{n-1}\bar{\psi}_{2}  ,\quad c^{\dagger} \bar{\psi}_{-3n-3} =  (48 -24 n (-7 + 9 n^2)) \bar{\psi}_{-3(n-1)-3} \]
\[ c^{\dagger} \psi_{2-3n}  =-24 (n-1) ( 3 n-8) (3 n-4)\bar{\psi}_{2-3(n-1)}  ,\quad c^{\dagger} \bar{\psi}_{-3n-3} =  (48 -24 n (-7 + 9 n^2)) \bar{\psi}_{-3(n-1)-3} \]
\[ c^{\dagger} \bar{\psi}_{3n-3} = \bar{\psi}_{3(n+1)-3}  ,\quad c^{\dagger} \psi_{3n-3} = \psi_{3(n+1)-3} \, . \]

We have now presented a complete description of the 2 chain representations in an algebraic manner. Those results demonstrate how all the states are, in fact, obtained via a minimal set of states, both polynomial and non-polynomial. More importantly, they transform the problem of obtaining the coefficients from a problem of higher-order differential operators acting on a non-trivial solution of the EOP Hermite ODE into a simpler problem where the commutation relations can be determined and used to build all the coefficients explicitly as third-degree polynomials in terms of $n$.

\par
%
%
\section{Conclusion}

In this paper we obtain explicit expressions for 2-chain representations for the polynomial Heisenberg algebra of the deformed oscillator related to $X_2$. This demonstrates how the polynomial, but also the non-polynomial part, play a role from a mathematical perspective. We build those 2-chain representations directly by constructing the states on $\mathbb{Z}$ and by then constructing the linearly independent solutions using the connection with the harmonic oscillator ( Hermite ) ODE. Non-trivial coefficients are then obtained for the action of the standard and non-standard ladder operators, both third-order differential operators. \par
%
%
We present an alternative point of view by using recently introduced general solutions \cite{cha20} and point out how those special functions beyond the hypergeometric are suitable for a description of those representations. We also discuss the construction in the context of the confluent form of the equation and also point out how the results can be used to build ladders for the Heun equation.
\par
%
%
One of the main results of the paper is the construction of a Rodrigues analogue which allows us to construct all the states in an algebraic way. We then use commutator identities to calculate the action of lowering and raising operators in a purely algebraic way. All those results describe a ladder for the EOP $X_2^{(1)}$ Hermite ODE for the general solution from a different perspective related to the theory of ODE and of algebraic structures beyond Lie algebras. Those 2-chain representations are also interesting as they extend finite and infinite dimensional 1-chain representations made by the polynomial part of the solution. The connection of EOP Hermite equations \cite{mar16,zel22} and Heun equation \cite{der22} with Painlev\'e transcendents were the objectives of several works. Those connection may allow us to understand equation of the Painlev\'e transcendents and their properties.
\par
%
%
.\par
%
%
\section*{Acknowledgments}

AMG was supported by a research grant from NSERC of Canada. DL was supported by Australian Research Council Discovery Project DP190101529 (A/Prof. Y.-Z. Zhang). IM was supported by the Australian Research Council Fellowship FT180100099. I.M. thanks the Centre de Recherches Math\'ematiques, Universit\'e de Montr\'eal for its hospitality.  \par
%
%

\par
%
%

\end{document}